\documentclass[12pt]{iopart}

\usepackage{graphicx}  
\usepackage{longtable} 
\usepackage{caption} 
\begin{document}

\title[Capitalizing on Nuclear Data Libraries' Comprehensiveness ....]{Capitalizing on Nuclear Data Libraries' Comprehensiveness to Obtain Solar System \mbox{$r$-process} Abundances}

\author{Boris Pritychenko}

\address{National Nuclear Data Center, Brookhaven National Laboratory, \\ Upton, NY 11973-5000, USA}
\ead{pritychenko@bnl.gov}
\vspace{10pt}
\begin{indented}
\item[]June 18, 2021
\end{indented}

\begin{abstract}
The recent observation of neutron stars merger by the Laser Interferometer Gravitational-Wave Observatory (LIGO) collaboration and the measurements of the event's electromagnetic spectrum as a function of time for different wavelengths have altered profoundly our understanding of the \mbox{$r$-process} site as well as considerably energized nuclear astrophysics research efforts. \mbox{$R$-process} abundances are a key element in \mbox{$r$-process} simulations, as a successful calculation must account for these abundances in the final debris of a stellar cataclysmic event. In this article, mankind's complete knowledge of neutron cross sections obtained over the last 80 years, as encapsulated in the latest release of the Evaluated Nuclear Data File (ENDF/B) library, is used to obtain solar system \mbox{$r$-process} abundances in a comprehensive data approach. ENDF/B cross sections have been successfully used for decades in nuclear power and defense applications and are now used to obtain \mbox{$r$-process} abundances in a fully traceable and documented  way.    This article \mbox{$r$-process} abundances provide complementary insights on the astrophysical events, overall quality of neutron capture data in the astrophysical region of temperatures, and demonstrate issues with  the  \mbox{$s$-process} contribution subtraction procedure. 

\end{abstract}

%
%
%
%
%

Understanding the origin of the elements has been a long intellectual adventure~\cite{19Woo}.  A scientifically sound theory began to materialize by the late 1950s, thanks to an emerging wealth of nuclear physics, chemistry, and solar system abundance data~\cite{57Bur,57Cam}.   In particular, it was understood that most of the heavy elements could only be produced by neutron capture following two possible mechanisms:   (1) the slow capture or $s$-process, characterized by low densities and temperatures environments and (2) the rapid capture or $r$-process, where  both densities and temperatures are large enough that successive neutron captures could lead in within seconds  to the synthesis of Uranium nuclides and possibly beyond.    It is currently thought that Asymptotic Giant Branch and red giant stars are the site for the $s$-process; while neutron star mergers, core-collapsed supernovae and magneto-rotational (MHD-jet) magnetars would produce the density and temperature needed for the $r$-process~\cite{74Lat,17Thi}.  The recent multi-messenger observation of a binary neutron stars merger (GW170817) and the indirect clues of \mbox{$r$-process} production of lanthanide elements~\cite{17Sma,17Tan} have further refined our understanding of the \mbox{$r$-process} site and exemplified the benefits of synergistic observations.  The lanthanides production findings are based on atomic physics models and include additional uncertainties. 
 
Several sets of data needed to understand the \mbox{$r$-process}, including the sun and  metal-poor stars \mbox{$r$-process} abundances.  The solar system $r$-process abundances are tabulated as a function of atomic mass (A) for the stable nuclides produced  as residual after subtracting the \mbox{$s$-process} and, in some cases,  proton captures from the total solar system abundances. The low-metallicity older stars elemental abundances are based on atomic spectra observations, and they provide a window into early nucleosynthesis~\cite{14Roe,09Sne,16Ji}.  The obtention of a set of solar \mbox{$r$-process} abundances is the core of this article and will be extensively discussed in the following paragraphs.  
 
The solar system residuals include 21 stable nuclei solely produced by the \mbox{$r$-process}  in the 100 $\leq$ A $\leq$ 208 range and a large number of nuclei with the \mbox{$s$-process} contributions.  Knowledge of the \mbox{$r$-process-only} nuclei is not sufficient to obtain a comprehensive picture of the \mbox{$r$-process} production pattern; one must subtract the \mbox{$s$-process} component from nuclides that can be produced by both processes.   For instance, in the well-known \mbox{$r$-process} peak in the 190 $\leq$ A $\leq$ 200 range caused by the $N$=126 magic number, only two nuclides, $^{192}$Os and $^{198}$Pt, are \mbox{$r$-process-only} nuclides; the peak truly emerges when subtracting the \mbox{$s$-process} contribution.  Previously, the Maxwellian-averaged cross sections (MACS) compiled by the Karlsruhe Astrophysical Database of Nucleosynthesis in Stars (KADoNiS)~\cite{06Dil} were extensively used in $s$- and \mbox{$r$-process} simulations.   Many of the KADoNiS MACS were normalized to a $^{197}$Au(n,$\gamma$) activation measurement that produced a 30-keV MACS equal to 582$\pm$9 mb~\cite{88Rat}.   For years, the disagreement between this value and that from the international evaluation of neutron cross-section standards, 620$\pm$11 mb~\cite{15Car,18Car}, was not understood.   However, it was recently resolved by re-analyzing activation measurements. The new KADoNiS gold value is 612$\pm$6 mb~\cite{18Rei}, which has led to an identification and  renormalization effort of 63 KADoNiS cross sections.  These relatively small astrophysical data sets deviations represent a serious problem for the stellar pattern analysis where precision up to 1$\%$ is needed~\cite{11FKa}.

The current analysis reveals a strong need for a reliable \mbox{$s$-process} model  to predict the \mbox{$s$-process} component using the solar system observables and nuclear physics inputs. The model should include high-fidelity neutron cross section data sets over the broad energy range with comprehensive coverage of neutron resonances.  Such data sets have been developed for nuclear power and national security applications  worldwide. In the USA, the Cross Section Evaluation Working Group (CSEWG) was created in 1966~\cite{CSEWG} and charged to produce the ENDF/B library.   The library was first released in 1968 with its latest release in 2018~\cite{18Bro}. This release included efforts from 70 people and 29 organizations worldwide.    The ENDF/B library contains recommended values of neutron-induced cross sections,  first compiled in the Experimental Nuclear Reaction Data (EXFOR) library~\cite{EXFOR}, then critically reviewed and augmented with extensive use of R-matrix and Hauser-Feschbach codes, and finally validated with a vast set of integral experiment benchmarks.  The recommended cross sections are tabulated in the laboratory system and include comprehensive covariance matrices obtained not only from experimental conditions but also by including nuclear physics model correlations and validated with integral experiments.  The ENDF/B library consists of unique reaction data evaluations that are  based on all available nuclear physics results evaluated and optimized across  the broad 10$^{-5}$ eV - 20 MeV energy span. The standalone ENDF/B-VIII.0 $^{197}$Au neutron capture evaluation incorporates the main neutron time of flight measurements ~\cite{18Rei,18Bro} that are essentially free of systematic errors associated with activation measurements. 
The ENDF/B library development benefits from CSEWG's cumulative and collective experience as well as by frequent interactions with other similar national and regional organizations. These include Japanese Evaluated Nuclear Library (JENDL)~\cite{11Shi}, Joint Evaluated Fission and Fusion (JEFF) European nuclear data library~\cite{11Kon}, Chinese Evaluated Nuclear Library (CENDL)~\cite{11Ge}, and the Russian Fund of Evaluated Neutron Data (ROSFOND)~\cite{07Zab}.  These exchanges are facilitated through the coordinating efforts of organizations, such as the Organisation for Economic Co-operation and Development (OECD) Nuclear Energy Agency Data Bank~\cite{OECD} and the International Atomic Energy Agency (IAEA) Nuclear Data Section~\cite{IAEA}.  The recently released  ENDF/B-VIII.0  library~\cite{18Bro}  includes many new evaluations, such as an \mbox{$s$-process} seed $^{56}$Fe~\cite{18Her}, and incorporates data from the sixth edition of the Atlas of Neutron Resonances~\cite{18Atlas}. Neutron resonances  dominate the astrophysical range of temperatures in many nuclei, and the comprehensive analysis of resonances is essential in nuclear astrophysics applications.

 In this article, the well-understood classical model of $s$-process nucleosynthesis~\cite{74Cla,82Kap} and the ENDF/B-VIII.0  library recommended cross sections~\cite{18Bro}  will be used to quantify the \mbox{$s$-process} abundance contributions.  The classical model is based on a phenomenological and site-independent approach, and it  assumes that the seeds for neutron captures are made entirely of $^{56}$Fe.  
The \mbox{$s$-process} abundance of an isotope $N_{(A)}$ depends on its precursor $N_{(A-1)}$ quantity as in
\begin{equation}\label{eq:class1}
\frac{dN_{(A)}}{dt} = \lambda_{n (A-1)} N_{(A-1)} - \big[ \lambda_{n (A)} + \lambda_{\beta (A)} \big] N_{(A)},
\end{equation}
where $\lambda_{n}$ is the neutron capture rate, and $\lambda_{\beta} = \frac{ln 2}{T_{1/2}}$ is the $\beta$-decay rate for radioactive nuclei. 
Assuming that the temperature and neutron density are constant,  and neglecting  \mbox{$s$-process} branchings, the previous formula simplifies to 
\begin{equation}\label{eq:class2}
\frac{dN_{(A)}}{dt} = \sigma_{(A-1)} N_{(A-1)} - \sigma_{(A)} N_{(A)}.
\end{equation}

Equation  \ref{eq:class2} was solved analytically  for an exponential average flow of neutron exposure 
assuming that  temperature remains constant over the whole timescale of the \mbox{$s$-process}~\cite{74Cla,82Kap}. The product of MACS and isotopic abundance ($\sigma_{(A)} N_{(A)}$) was written as
\begin{equation}\label{eq:class3}
\sigma_{(A)} N_{(A)} = \frac{f N_{56}}{\tau_{0}} \prod_{i=56}^{A} \big[ 1+ \frac{1}{\sigma(i) \tau_{0}} \big]^{-1},
\end{equation}
where  $f$ and $\tau_{0}$ are the neutron fluence distribution parameters, and $N_{56}$ is the initial abundance of $^{56}$Fe seed. 

Center of mass system MACS  in Eq. \ref{eq:class3} are described as   
\begin{equation}
\label{myeq.max3}
\sigma^{Maxw}(kT) = \frac{2}{\sqrt{\pi}} \frac{a^{2}}{(kT)^{2}}  \int_{0}^{\infty} \sigma(E^{L}_{n})E^{L}_{n} e^{- \frac{aE^{L}_{n}}{kT}} dE^{L}_{n},
\end{equation}
where $a = m_2/(m_1 + m_2)$, {\it k} and {\it T} are the Boltzmann constant and temperature of the system, respectively,  and $E$ is the energy of relative motion of 
the neutron with respect to the target. $E^{L}_{n}$ is the neutron energy in the laboratory system,  while $m_{1}$ and $m_{2}$ are the masses of 
the neutron and the target nucleus, respectively~\cite{92Beer,05Nak}. Equation \ref{myeq.max3} has been used in the present work to  calculate ENDF/B-VIII.0 MACS at $kT$=30 keV.  Prior to these calculations, the  neutron resonance region evaluated data had been Doppler broadened, assuming a target temperature of T=293.16 K with the PREPRO code~\cite{Prepro}  and the room-temperature cross sections were complemented with  the stellar enhancement factors (SEF)  of Rauscher~\cite{09Rau}.

 Analysis of pre-solar samples~\cite{09Lod,07Arn} shows that \mbox{$s$-process} abundances originate from a superposition of the
two major exponential distributions of time-integrated neutron exposure: (1) weak component (responsible for the production of \mbox{70 $\leq$ A $\leq$ 90 nuclei})  
and (2) the main component (for \mbox{90 $\leq$ A $\leq$ 204 nuclei}).  Previously,  \mbox{$s$-process} experimental cross sections have been analyzed and fitted from $^{56}$Fe to $^{210}$Po  as a sum of  the two  components   that were individually described by \mbox{Eq.~\ref{eq:class3} of Ref.~\cite{82Kap}}. Herein   cross sections and solar system abundances were taken from the presently outdated compilations~\cite{82Kap,81Cam} and optimized for \mbox{$s$-process-only} target nuclei.  In the fit of a weak \mbox{$s$-process} component, K{\"a}ppeler  et al. have included $^{88}$Sr, $^{89}$Y, and $^{90}$Zr to overcome a relatively small number of \mbox{$s$-process-only} nuclei in the $A<90$ region. K{\"a}ppeler  et al. have argued that the above-mentioned nuclei solar system abundances have $<20 \%$ \mbox{$r$-process} contributions, and they could be used in the fitting process.  An attempt to reproduce the two-component fitting  using the  present-day cross sections and abundances  was not successful. The present findings are consistent with  Arlandini et al.~\cite{99Arl}, who used the classical model to calculate the \mbox{$s$-process} main component and extract \mbox{$r$-process} abundances. Arlandini et al. assumed that between Cu and Sr, the contribution of the weak s-component is described by the single-exposure calculation of Beer et al. \cite{92Bee}. Subsequent analysis of K{\"a}ppeler et al.~\cite{11Kap} shows that the weak component is not firmly described by the classical analysis because of a limited number of \mbox{$s$-process-only} medium nuclei  and lack of equilibrium conditions. The empirical $\sigma N$ values for heavy nuclei that are not affected by branchings are reproduced with a mean square deviation of only 3$\%$. 

Therefore, the main component only findings are shown in the present work and nuclei abundances are taken from Ref.~\cite{09Lod}.   Neutron fluence parameters for \mbox{$s$-process}-only isotopes  were derived using  Eq. \ref{eq:class3} above.  Later, the derived parameters were optimized using  least squares procedures, and  $f$ and $\tau_{0}$  neutron fluence distribution numerical values were obtained. The resulting fluence parameters  are shown in Table~\ref{tab:sfits}.

\begin{table}[h!]
\begin{center}
 \caption{$S$-process main component neutron fluence distribution parameters for ENDF/B-VIII.0  library~\cite{18Bro}.  \label{tab:sfits} 
 }

\begin{tabular}{l|c}
     \hline\hline
Parameters	&	ENDF/B-VIII.0  \\  \hline
f		&	0.000402$\pm$0.000083			\\
$\tau_{0}$	&	0.332365$\pm$0.027410			\\    \hline\hline
  \end{tabular}
\end{center}
\end{table}

\begin{figure}
\begin{center}
\includegraphics[width=.65\linewidth]{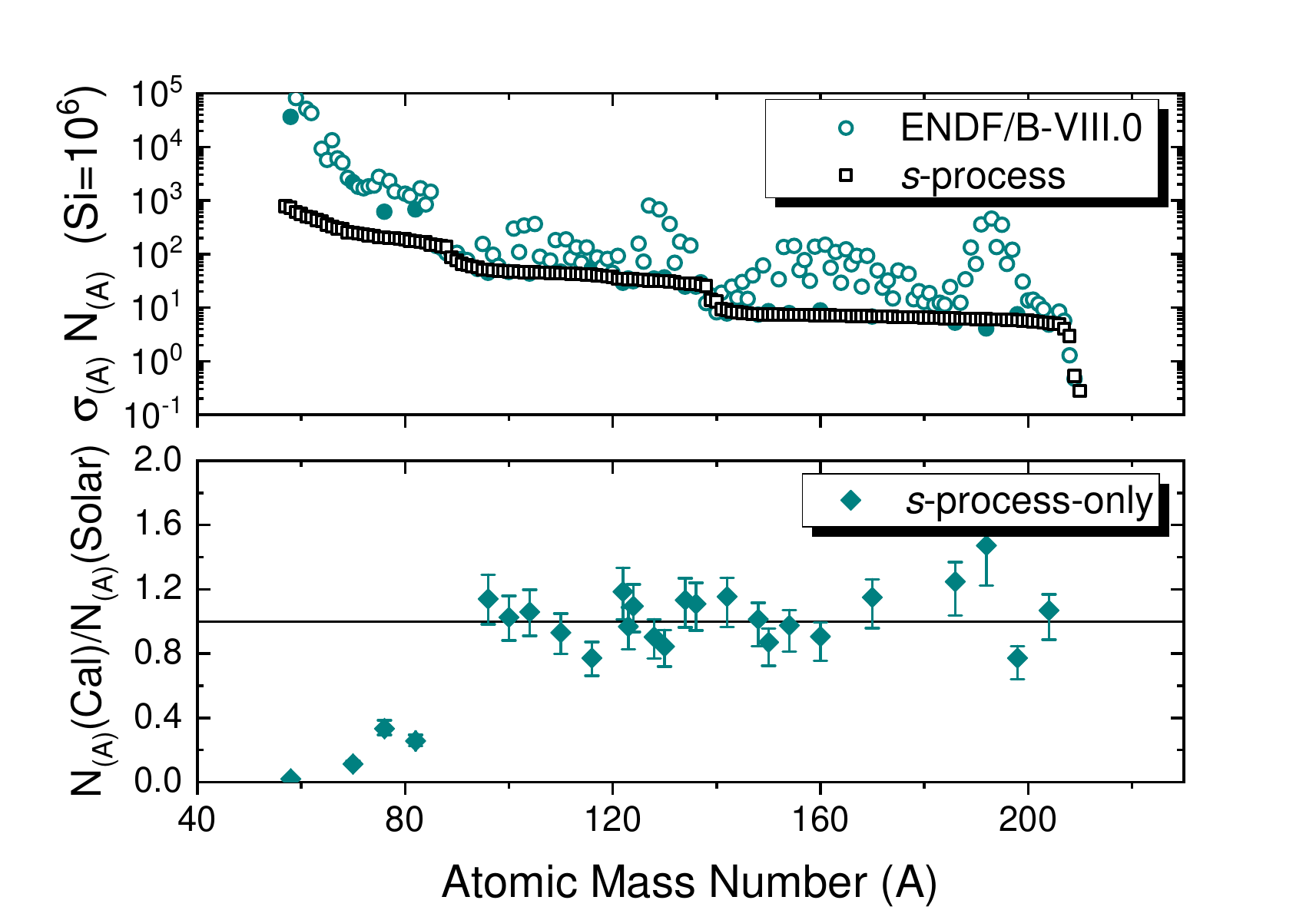}
\end{center}
\caption{ 
Upper panel: ENDF/B-VIII.0 (n,$\gamma$) MACS at 30 keV times solar system abundances (circles) as function of the atomic mass number for $s$- and \mbox{$r$-process} nuclides.  The \mbox{$s$-process-only} nuclei are shown as full circles, and the squares correspond to the classical \mbox{$s$-process} values derived as described in the text. 
Lower panel: Ratio of the calculated to the solar system abundances for  $s$-process-only nuclei. 
\label{fig:snuclei}}
\end{figure}

The \mbox{$s$-process} contribution to solar system abundances can be estimated using  neutron fluence parameters and compared with 
 observed values. The ENDF/B-VIII.0 MACS at $kT$=30 keV times abundance and expected classical model product values are shown in the upper panel of Fig. \ref{fig:snuclei}. The calculated ratios for $s$-process-only nuclei are shown in the lower panel of Fig. \ref{fig:snuclei}.

\begin{figure}
\begin{center}
 \includegraphics[width=.65\linewidth]{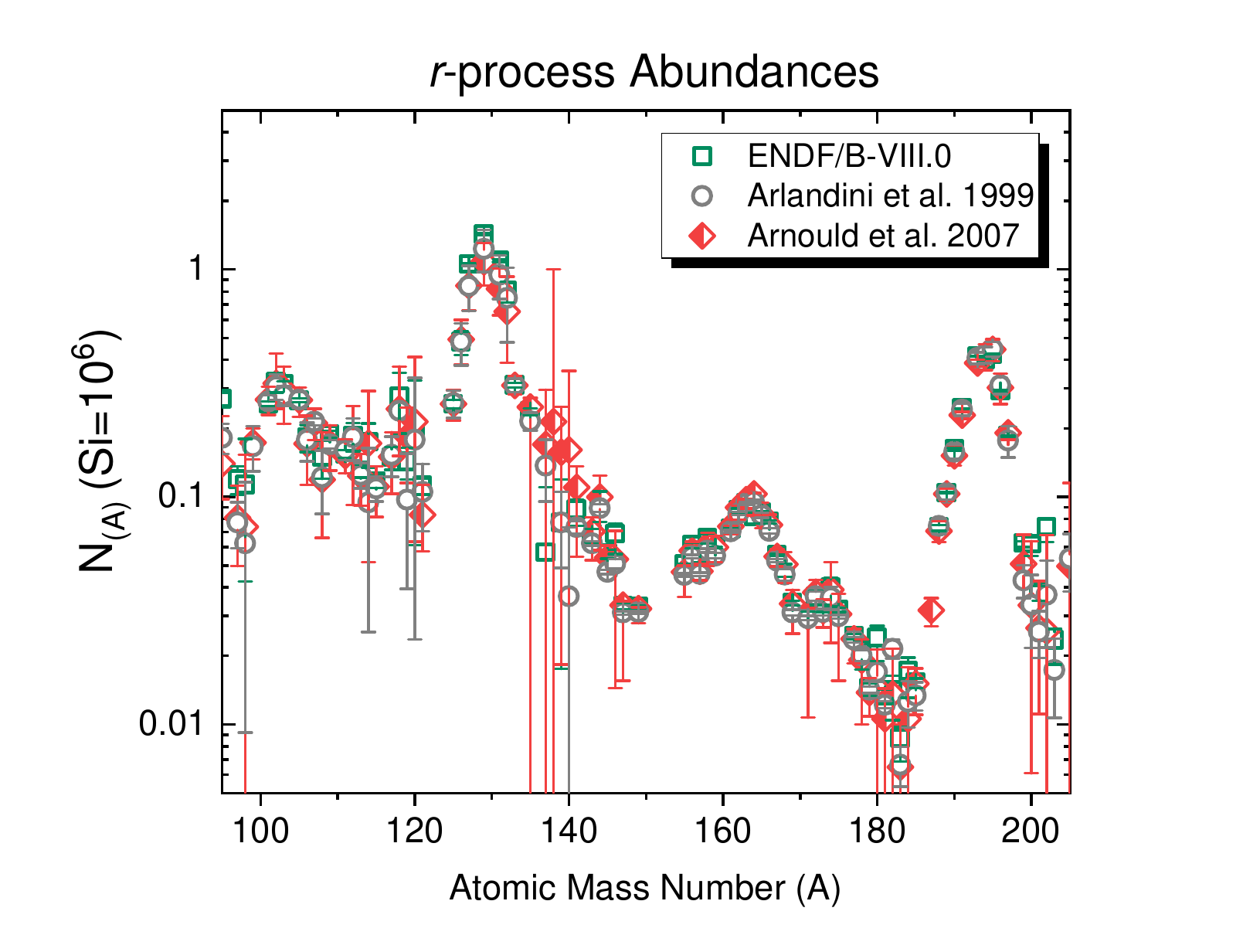}
\end{center}
\caption{ 
 Solar \mbox{$r$-process} abundances for nuclides that are produced by both the $s$- and \mbox{$r$-processes} derived from ENDF/B-VIII.0 (squares) compared with those obtained by Arlandini et al.~\cite{99Arl} (circles) and Arnould et al.~\cite{99Gor,07Arn,93Pal} (diamonds). 
} \label{fig:rnuclei2}
\end{figure}

The data in the figure indicate a surplus production for many nuclei compared with the \mbox{$s$-process} expectations. 
This surplus is commonly attributed to the \mbox{$r$-process} contribution, and it can be obtained  by subtracting the 
expected classical model \mbox{$s$-process} production from the total values. 
     The  classical model \mbox{$r$-process} abundances derived from ENDF/B-VIII.0  library  and  Arlandini et al. \cite{99Arl} together with a multi-component model fits of Goriely~\cite{99Gor} $\&$ Arnould et al.~\cite{07Arn} are shown in Fig.~\ref {fig:rnuclei2} and Table~\ref{RTable102}.  The present $r$-process abundances are based on the $s$-process main component calculation only. The main component complements the weak $s$-process component production of medium nuclei, and it is dominant for heavy nuclei. These complementary values were used in the present work to extract the upper bounds on $r$-process abundances of medium nuclei from Ga to Sr.   For comparative purposes, the \mbox{$s$-process} main component values calculated by Arlandini et al.~\cite{99Arl} were deduced using the Anders and Grevesse~\cite{89And} solar system abundances in $Z=31-38$ region.     
     
     The analysis of Goriely Table 3~\cite{99Gor} and Arnould  et al. Table 1 data~\cite{07Arn} shows multiple coincidences, it was assumed that Goriely values were superseded by Arnould et al.      The numerical values and comments for the present work,  Arlandini et al.~\cite{99Arl} and Arnould et al.~\cite{07Arn} are given in the Appendix. 
The  present work \mbox{$r$-process} uncertainties are due to  least squares fitting of \mbox{$s$-process-only} nuclei $\sigma N$ product  and subtraction of the classical model contribution from the total product values. The total product value uncertainties are solely based on ENDF/B cross section uncertainties since the solar system abundances of Lodders, Palme and Gail~\cite{09Lod} contain absolute values only, and stellar enhancement factors are calculated from nuclear theory~\cite{09Rau}.  The Fig.~\ref{fig:rnuclei2}  data analysis shows the second and third \mbox{$r$-process} abundance peaks and the broad surge due to production of lanthanides that were tentatively found in neutron stars merger~\cite{17Sma,17Tan}.  These peaks provide indisputable evidence that nuclear shell closing (magic numbers) persists outside the valley of stability for $N \sim$ 82 and 126.
The current \mbox{$r$-process}  abundances    agree well with  the classical analysis of \mbox{Arlandini et al. \cite{99Arl}} and values of Goriely~\cite{99Gor}  $\&$ Arnould et al.~\cite{07Arn} deduced from Ref.~\cite{93Pal} with an exception to $N$=82 lanthanide nuclei  where the current work data are not smooth and show structure. This observation concurs with Kratz et al.~\cite{07Kra} who noticed that the subtraction of \mbox{$s$-process} contribution from the total values is accurate for nuclei with little \mbox{$s$-process} contribution but results in significant uncertainties when \mbox{$s$-process} fraction is dominant.  The issues with spectra subtraction are not unique to nuclear astrophysics; these phenomena are well documented in  nuclear reaction physics~\cite{75Go} and discoveries of double-beta decay in $^{76}$Ge and $^{100}$Mo~\cite{90Va,90Vas}. 

\begin{figure}
\begin{center}
\includegraphics[width=.65\linewidth]{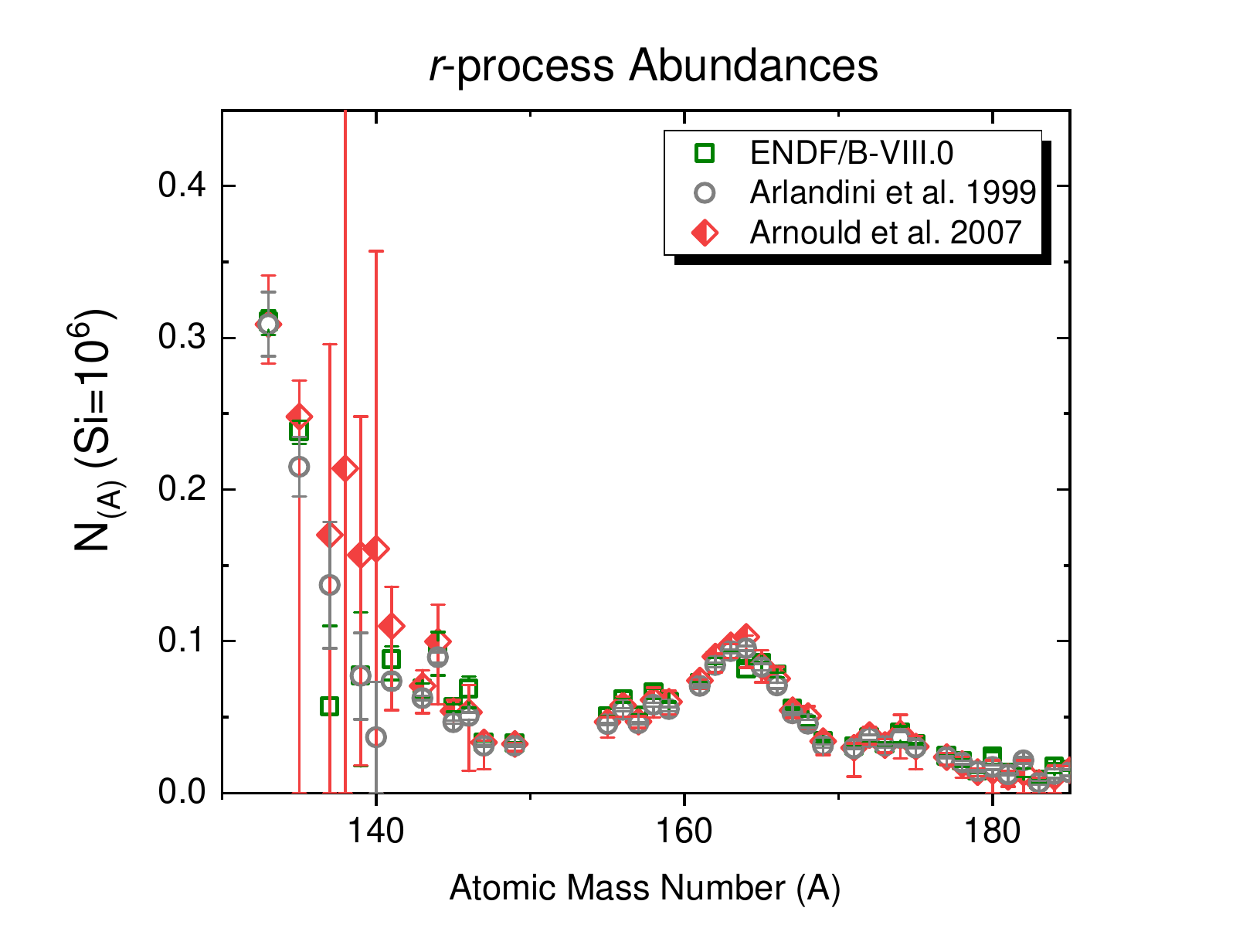}
\end{center}
\caption{Zoomed view of Fig. 2 that allows a better inspection of solar \mbox{$r$-process} abundances for the lanthanide nuclides (A=139-176). 
} \label{fig:rlanthan}
\end{figure}

The present work classical modeling ignores \mbox{$s$-process} branching nuclei; for calculation of these nuclei additional information on stellar neutron fluxes (exposures) and  accurate beta decay rates are needed. The stellar neutron fluxes are astrophysical site dependent while beta decay rates along the valley of stability in the Evaluated  Nuclear Structure Data File (ENSDF)~\cite{ensdf} are often impacted by relatively old and not precise measurements. In light of this disclosure, only the lower bounds for \mbox{$r$-process} abundances for branching nuclei are deduced in the present work.
Further examination of the lanthanide region abundances shown in Fig.~\ref{fig:rlanthan} demonstrates the high quality of ENDF/B-VIII.0 library \mbox{$r$-process} abundances with an exception to $^{138}$Ba and $^{140}$Ce  where the expected \mbox{$s$-process} contributions exceed total product values. The main \mbox{$s$-process} component overproduction in $^{138}$Ba has been previously reported by Palme $\&$ Beer~\cite{93Pal} and interpreted by Arnould et al.~\cite{07Arn} as \mbox{$r$-process} abundance of 0.214$^{+0.786}_{-0.214}$ (Si=10$^6$).  Arlandini et al.~\cite{99Arl} values for $^{138}$Ba and $^{140}$Ce are slightly below and above zero, respectively.  ENDF barium and cerium neutron capture cross sections are based on the SubGroup 23 (International Library of Fission Product Evaluations) recommendations~\cite{SG23}  clearly have to be revisited.  
These ENDF/B-VIII.0 MACS  solar system abundance~\cite{18Bro,09Lod} product deficiencies highlight mutually beneficial relations between nuclear data and astrophysics efforts, and they will be addressed in the next release of ENDF/B library.  The present results demonstrate a large potential of evaluated libraries for stellar nucleosynthesis calculations, the sensitivity studies of the impact of individual nuclear properties on \mbox{$r$-process} nucleosynthesis~\cite{16Mum}, and  analysis of  astrophysical observables. 

\begin{figure}
\begin{center}
\includegraphics[width=.65\linewidth]{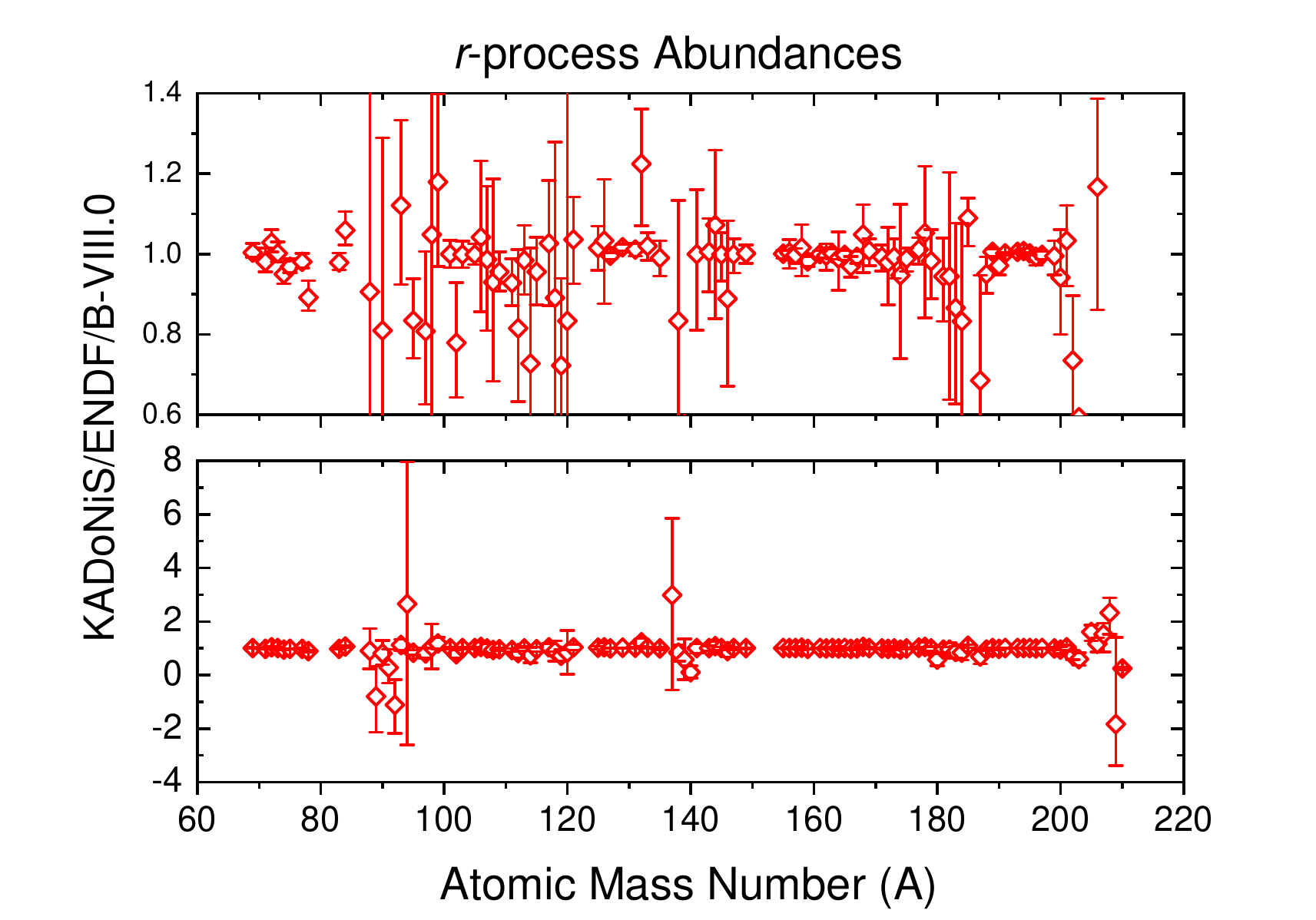}
\end{center}
\caption{ 
 The ratio  of KADoNiS 0.3 \cite{06Dil} to ENDF/B-VIII.0 solar \mbox{$r$-process} abundances.  Upper panel: Zoomed view. Lower panel: Complete view. The imperfectly subtracted residuals or \mbox{$s$-process} overproduction cases are depicted  as negative ratios. } \label{fig:K03E8}
\end{figure}

To evaluate the possible impact of different data sets, the  KADoNiS 0.3 library~\cite{06Dil} cross sections were processed in the current work, and the corresponding  \mbox{$r$-process} abundances were extracted. Figure~\ref{fig:K03E8} shows the KADoNiS 0.3 to ENDF/B-VIII.0 ratio of \mbox{$r$-process} abundances. The calculated abundances ratios include several outliers  near the neutron magic numbers $N$=50, 82, 126 and smaller deviations along the \mbox{$s$-process} path. Large discrepancies could be explained by the  KADoNiS 0.3 library over reliance on a single measurement or cross section calculation.  The smaller deviations may suggest an additional $\sim$20-30$\%$ systematic error  for present day nuclear astrophysics calculations. Some of these smaller deviations could be explained by the impact of an erroneous $^{197}$Au Karlsruhe cross  section~\cite{88Rat} on the KADoNiS library while others are due to the lack of experimental data. Further analysis of the KaDONis 0.3 library  cross sections along the \mbox{$s$-process} path shows   that $\sim$54$\%$ targets are affected by   Ratynski and K{\"a}ppeler neutron flux monitor value~\cite{88Rat}, $\sim$10$\%$ are based on other Karlsruhe measurements, $\sim$7$\%$ are purely theoretical, and $\sim$29$\%$ reflect experimental findings of other authors. The approximately 50/50 mix between gold monitored and  other cross sections effectively randomizes the impact of overestimated neutron flux at Karlsruhe on \mbox{$r$-process} abundances.  
Due to space and time constraints,  the detailed analysis of nuclear data for more than 150  \mbox{$s$-process} nuclei is not given in the current work. It will be addressed in  subsequent publications.

In conclusion, the indirect observations of \mbox{$r$-process} elements in the neutron stars merger renewed interest in stellar nucleosynthesis 
calculations and the corresponding nuclear data. Recent re-analysis of KADoNiS library~\cite{18Rei} reveals multiple issues with the Karlsruhe data and a strong need for complementary data sets. The release of the ENDF/B-VIII.0  library  creates a unique opportunity for nuclear science and technology developments and further exploration in nuclear astrophysics. The Maxwellian-averaged (n,$\gamma$) cross sections for 553 ENDF/B-VIII.0 library target nuclides have been produced. These data were combined with the solar system abundances and fitted. Astrophysical \mbox{$r$-process}  abundances have been extracted in the present work, compared with  available values, and an  agreement was deduced.

The next stage of the current project will involve incorporation of the evaluated nuclear data libraries into astrophysical model codes. 
Work on ENDF/B-VIII.0  library reaction rates data transfer for (n,$\gamma$), (n,$\alpha$), (n,p) and (n,fission) channels within 0.01-10 GK neutron temperatures into Reaction rate Library (REACLIB) format  \cite{00Rau,10Cyb,20Pri} is currently underway. The new data sets will provide a more extensive coverage of neutron resonances and will make REACLIB fits more reliable across the whole \mbox{$s$-process} temperature range of 8-90 keV \cite{11Kap}. 

\textit{Note added in proof:} The corrected values of Karlsruhe group cross sections are available in Ref.~\cite{21Kap}.

The author is indebted to A. Sonzogni for encouragement of this project,  D. Brown for useful comments, and  J. Frejka for careful reading of the manuscript and valuable suggestions.   Work at Brookhaven was funded by the Office of Nuclear Physics, Office of Science of the U.S. Department
of Energy, under Contract No. DE-SC0012704  with Brookhaven Science Associates, LLC. \\ \\


\newpage
\section*{Appendix}

The  present work of ENDF/B-VIII.0, Arlandini et al. \cite{99Arl} and Arnould et al. \cite{07Arn}  \mbox{$r$-process} abundances are shown in Table \ref{RTable102}.


\begin{footnotesize}

\begin{longtable}[h!]{@{\extracolsep\fill}p{1.9cm}|p{3.5cm}p{3.3cm}p{3.0cm}p{3.4cm}@{}}
\caption{{\it R}-process abundances (in the Si = 10$^{6}$ scale) obtained from ENDF/B-VIII.0  library~\cite{18Bro}, the  previous values of Arlandini et  al.~\cite{99Arl} and Arnould et al.~\cite{07Arn}.  \mbox{$R$-process-only} abundances for ENDF/B library are adopted from \mbox{4.56 Gy} ago values of Ref.~\cite{09Lod}; 
alpha and beta decay T$_{1/2}$ are taken from the Evaluated Nuclear Structure Data File~\cite{ensdf}. 
 The upper bounds on $r$-process abundances in the Z=31-38 region are marked with a $\star$ character, 
 while the lower bound for $s$-process branching nuclei are identified with a  $\dagger$ character.}\label{RTable102} \\ 
\hline\hline \\

 {\bfseries Target}  &    {\bfseries  ENDF Abundances} & {\bfseries Arlandini et al. \cite{99Arl}} & {\bfseries Arnould et al. \cite{07Arn}} & {\bfseries Comments} \\
 &     $N_{\odot}$-N$_{main}$ &  $N_{\odot}$-N$_{main}$   &  &  \\

\hline \\
\endfirsthead
\caption[]{{\it R}-process abundances  ...  (continued).} \\
\hline\hline \\
 {\bfseries Target} &    {\bfseries  ENDF Abundances} & {\bfseries Arlandini et al. \cite{99Arl}}  & {\bfseries Arnould et al. \cite{07Arn}} & {\bfseries Comments} \\
 &     $N_{\odot}$-N$_{main}$ &  $N_{\odot}$-N$_{main}$   &  &  \\
\hline \\
\endhead 

31-Ga-69	&	1.988E+1	$_{1.657E-1}^{3.533E-1}$$^{\star}$	&	1.985E+1$\pm$2.203E+0$^{\star}$	&	6.180E+0	$_{6.180E+0}^{3.190E+0}$	&	\\ 
32-Ge-70	&	2.155E+1	$_{2.161E-1}^{4.562E-1}$$^{\star}$	&	2.003E+1$\pm$2.243E+0$^{\star}$	&							&	$s$-process only		\\
30-Zn-70	&	8.000E+0						&	7.793E+0$\pm$3.912E+0	&	7.740E+0$^{\star}$	$_{9.400E-1}^{8.100E-1}$	&	$r$-process only		\\
31-Ga-71	&	1.266E+1	$_{2.163E-1}^{3.200E-1}$$^{\star}$	&	1.085E+1$\pm$9.765E-1$^{\star}$	&	1.960E+0	$_{1.960E+0}^{7.650E+0}$	&			\\
32-Ge-72	&	2.733E+1	$_{4.898E-1}^{7.188E-1}$$^{\star}$	&	2.632E+1$\pm$7.817E+0$^{\star}$	&	0.000E+0	$_{0.000E+0}^{9.930E+0}$	&			\\
32-Ge-73	&	7.756E+0	$_{1.182E-1}^{1.704E-1}$$^{\star}$	&	8.030E+0$\pm$2.473E+0$^{\star}$	&	6.310E+0	$_{6.310E+0}^{1.880E+0}$	&			\\
32-Ge-74	&	3.645E+1	$_{5.393E-1}^{7.735E-1}$$^{\star}$	&	3.729E+1$\pm$5.482E+0$^{\star}$	&	1.970E+1	$_{9.760E+0}^{9.200E+0}$	&			\\
33-As-75	&	5.650E+0	$_{5.160E-2}^{7.249E-2}$$^{\star}$	&	5.976E+0$\pm$7.530E-1$^{\star}$	&	3.780E+0	$_{5.400E-1}^{9.000E-1}$	&			\\
34-Se-76	&	4.221E+0	$_{2.413E-1}^{3.382E-1}$$^{\star}$	&	3.390E+0$\pm$2.712E-1$^{\star}$	&							&	$s$-process only		\\
32-Ge-76	&	8.500E+0						&	9.277E+0$\pm$1.169E+0$^{\star}$	&	8.780E+0	$_{9.400E-1}^{9.000E-1}$	&	$r$-process only		\\
34-Se-77	&	4.712E+0	$_{5.062E-2}^{7.024E-2}$$^{\star}$	&	3.973E+0$\pm$1.200E+0$^{\star}$	&	3.760E+0	$_{2.800E-1}^{8.900E-1}$	&			\\
34-Se-78	&	1.390E+1	$_{2.471E-1}^{3.421E-1}$$^{\star}$	&	1.126E+1$\pm$2.365E+0$^{\star}$	&	0.000E+0	$_{0.000E+0}^{1.030E+1}$	&			\\
34-Se-79/ 35-Br-79	&	-4.520E-1$_{5.257E-2}^{7.202E-2}$$^{\star}$$^{\dagger}$	&	5.022E+0$\pm$9.743E-1$^{\star}$	&	4.810E+0	$_{3.892E+0}^{9.000E-1}$	&	$s$-process branching;	$\beta$-, T$_{1/2}$=3.26x10$^{5}$ y	\\
34-Se-80	& 2.875E+1$_{5.509E-1}^{7.529E-1}$$^{\star}$$^{\dagger}$		&	2.741E+1$\pm$2.577E+0$^{\star}$	&	2.810E+1	$_{3.300E+0}^{4.100E+0}$	&	$s$-process branching		\\
34-Se-81/ 35-Br-81	& 4.524E+0$_{8.912E-2}^{1.190E-1}$$^{\star}$$^{\dagger}$		&	5.256E+0$\pm$1.020E+0$^{\star}$	&	4.070E+0	$_{1.030E+0}^{8.000E-1}$	&	$s$-process branching;	$\beta$-, T$_{1/2}$=18.45 min	\\
36-Kr-82	&	4.850E+0	$_{1.962E-1}^{2.609E-1}$$^{\star}$	&	1.770E+0$\pm$3.469E-1$^{\star}$	&							&	$s$-process only		\\
34-Se-82	&	5.890E+0						&	5.698E+0$\pm$2.872E+0$^{\star}$	&	6.200E+0	$_{3.700E-1}^{3.100E-1}$	&	$r$-process only		\\
36-Kr-83	&	5.829E+0	$_{7.370E-2}^{9.710E-2}$$^{\star}$	&	4.040E+0$\pm$7.716E-1$^{\star}$	&	4.380E+0	$_{1.330E+0}^{1.300E+0}$	&			\\
36-Kr-84	&	2.557E+1	$_{7.387E-1}^{9.696E-1}$$^{\star}$	&	1.749E+1$\pm$3.760E+0$^{\star}$	&	2.360E+1	$_{9.400E+0}^{1.090E+1}$	&			\\
36-Kr-85/ 37-Rb-85	& 4.599E+0$_{6.315E-2}^{8.021E-2}$$^{\star}$$^{\dagger}$		&	2.980E+0$\pm$2.265E-1$^{\star}$	&	2.870E+0	$_{1.820E+0}^{1.140E+0}$	&	$s$-process branching;	$\beta$-,  T$_{1/2}$=10.739 y	\\
38-Sr-86	& -6.727E-2$_{2.869E-1}^{3.632E-1}$$^{\star}$$^{\dagger}$		&	7.400E-1$\pm$6.512E-2$^{\star}$	&							&	$s$-process branching\\ 
38-Sr-87	& -1.226E-1$_{2.103E-1}^{2.624E-1}$$^{\star}$$^{\dagger}$		&	3.900E-1$\pm$3.471E-2$^{\star}$	&	2.920E-1	$_{2.920E-1}^{7.180E-1}$	&	$s$-process branching		\\
38-Sr-88	&	-5.704E+0	$_{3.059E+0}^{3.771E+0}$$^{\star}$	&	1.200E+0$\pm$9.720E-2$^{\star}$	&	4.090E+0	$_{4.090E+0}^{6.600E-1}$	&			\\
39-Y-89	&	6.187E-1	$_{5.197E-1}^{5.746E-1}$	&	-2.970E-1$\pm$-1.960E-2	&	1.110E+0	$_{1.110E+0}^{7.000E-1}$	&			\\
40-Zr-90	&	1.530E+0	$_{5.295E-1}^{5.641E-1}$	&	1.860E+0$\pm$2.306E-1	&	2.610E+0	$_{1.350E+0}^{4.000E-1}$	&			\\
40-Zr-91	&	2.304E-1	$_{1.316E-1}^{1.346E-1}$	&	-6.400E-3$\pm$-9.472E-4	&	2.100E-1	$_{2.100E-1}^{2.740E-1}$	&			\\
40-Zr-92	&	3.536E-1	$_{2.020E-1}^{2.039E-1}$	&	-1.627E-1$\pm$-2.229E-2	&	6.200E-2	$_{6.200E-2}^{3.750E-1}$	&			\\
40-Zr-93/41-Nb-93	& -5.724E-1$_{7.806E-2}^{7.725E-2}$$^{\dagger}$		&	-1.396E-2$\pm$-3.350E-4	&	9.870E-2	$_{9.870E-2}^{1.713E-1}$	&	$s$-process branching;	$\beta$-, T$_{1/2}$=1.61x10$^{6}$ y 	\\
40-Zr-94	& -1.477E-1$_{2.766E-1}^{2.714E-1}$$^{\dagger}$		&	-3.168E-1$\pm$-2.313E-2	&	0.000E+0	$_{0.000E+0}^{6.020E-2}$	& $s$-process branching			\\
40-Zr-95/ 41-Nb-95/ 42-Mo-95	&	2.695E-1	$_{1.865E-2}^{1.779E-2}$	&	1.820E-1$\pm$1.256E-2	&	1.400E-1	$_{4.240E-2}^{8.600E-2}$	&		$\beta$-, T$_{1/2}$=64.032 d; $\beta$-, T$_{1/2}$=34.991 d	\\
42-Mo-96	&	-5.911E-2	$_{6.718E-2}^{6.393E-2}$	&	-6.800E-2$\pm$-4.828E-3	&							&	$s$-process only		\\
40-Zr-96	&	3.020E-1						&	1.570E-1$\pm$1.146E-2	&	0.000E+0	$_{0.000E+0}^{2.500E-2}$	&	$r$-process only		\\
42-Mo-97	&	1.197E-1	$_{1.746E-2}^{1.648E-2}$	&	7.700E-2$\pm$5.313E-3	&	8.080E-2	$_{3.120E-2}^{3.120E-2}$	&			\\
42-Mo-98	&	1.134E-1	$_{7.091E-2}^{6.678E-2}$	&	6.300E-2$\pm$4.725E-3	&	7.390E-2	$_{7.390E-2}^{7.910E-2}$	&			\\
42-Mo-99/ 43-Tc-99/ 44-Ru-99	&  -4.381E-2$_{6.131E-3}^{5.723E-3}$$^{\dagger}$		&	1.666E-1$\pm$8.380E-2	&	1.730E-1	$_{2.700E-2}^{2.700E-2}$	& $s$-process branching;		$\beta$-, T$_{1/2}$=65.976 h; $\beta$-,  T$_{1/2}$=2.111x10$^{5}$ y	\\
44-Ru-100	&	-5.553E-3	$_{3.214E-2}^{2.997E-2}$	&	-2.574E-2$\pm$-2.136E-3	&							&	$s$-process only		\\
42-Mo-100	&	2.500E-1						&	2.460E-1$\pm$5.240E-2	&	2.260E-1	$_{1.600E-2}^{2.400E-2}$	&	$r$-process only		\\
44-Ru-101	&	2.566E-1	$_{6.646E-3}^{6.173E-3}$	&	2.622E-1$\pm$1.757E-2	&	2.670E-1	$_{3.700E-2}^{3.800E-2}$	&			\\
44-Ru-102	&	3.208E-1	$_{3.384E-2}^{3.140E-2}$	&	3.060E-1$\pm$2.448E-2	&	3.150E-1	$_{7.100E-2}^{1.110E-1}$	&			\\
45-Rh-103	&	3.136E-1	$_{7.935E-3}^{7.331E-3}$	&	2.790E-1$\pm$2.288E-2	&	2.970E-1	$_{8.800E-2}^{7.800E-2}$	&			\\
46-Pd-104	&	-9.022E-3	$_{2.255E-2}^{2.081E-2}$	&	-2.170E-2$\pm$-2.604E-3	&							&	$s$-process only		\\
44-Ru-104	&	3.320E-1						&	3.443E-1$\pm$2.823E-2	&	3.370E-1	$_{3.900E-2}^{4.600E-2}$	&	$r$-process only		\\
46-Pd-105	&	2.656E-1	$_{5.297E-3}^{4.874E-3}$	&	2.669E-1$\pm$2.215E-2	&	2.660E-1	$_{4.200E-2}^{3.700E-2}$	&			\\
46-Pd-106	&	1.834E-1	$_{2.643E-2}^{2.430E-2}$	&	1.780E-1$\pm$2.118E-2	&	1.710E-1	$_{5.800E-2}^{5.700E-2}$	&			\\
46-Pd-107/ 47-Ag-107	& -3.386E-2$_{4.777E-3}^{4.377E-3}$$^{\dagger}$		&	2.140E-1$\pm$8.988E-3	&	2.110E-1	$_{3.300E-2}^{3.300E-2}$	&	$s$-process branching; $\beta$-, T$_{1/2}$=6.5x10$^{6}$ y		\\
46-Pd-108	&  1.495E-1$_{2.957E-2}^{2.707E-2}$$^{\dagger}$		&	1.220E-1$\pm$1.452E-2	&	1.190E-1	$_{5.300E-2}^{7.400E-2}$	&	$s$-process branching		\\
47-Ag-109	&	1.886E-1	$_{6.708E-3}^{6.117E-3}$	&	1.696E-1$\pm$6.954E-3	&	1.720E-1	$_{4.100E-2}^{3.500E-2}$	&			\\
48-Cd-110	&	1.399E-2	$_{2.589E-2}^{2.358E-2}$	&	1.000E-3$\pm$1.380E-4	&							&	$s$-process only		\\
46-Pd-110	&	1.590E-1						&	1.627E-1$\pm$2.474E-2	&	1.560E-1	$_{2.000E-2}^{1.800E-2}$	&	$r$-process only		\\
48-Cd-111	&	1.550E-1	$_{6.510E-3}^{5.908E-3}$	&	1.599E-1$\pm$2.143E-2	&	1.520E-1	$_{2.500E-2}^{2.700E-2}$	&			\\
48-Cd-112	&	1.858E-1	$_{2.752E-2}^{2.495E-2}$	&	1.830E-1$\pm$2.635E-2	&	1.760E-1	$_{8.390E-2}^{7.400E-2}$	&			\\
48-Cd-113	&	1.308E-1	$_{8.685E-3}^{7.844E-3}$	&	1.247E-1$\pm$1.596E-2	&	1.240E-1	$_{3.240E-2}^{3.100E-2}$	&			\\
48-Cd-114	&	1.745E-1	$_{3.941E-2}^{3.555E-2}$	&	9.500E-2$\pm$1.596E-2	&	1.720E-1	$_{1.205E-1}^{1.190E-1}$	&			\\
49-In-115	&	1.172E-1	$_{7.518E-3}^{6.744E-3}$	&	1.094E-1$\pm$1.291E-2	&	1.110E-1	$_{2.940E-2}^{2.500E-2}$	&			\\
50-Sn-116	&	1.192E-1	$_{5.765E-2}^{5.166E-2}$	&	6.600E-2$\pm$6.270E-3	&							&	$s$-process only		\\
48-Cd-116	&	1.180E-1						&	1.118E-1$\pm$1.599E-2	&	9.550E-2	$_{2.580E-2}^{3.150E-2}$	&	$r$-process only		\\
50-Sn-117	&	1.495E-1	$_{1.822E-2}^{1.619E-2}$	&	1.530E-1$\pm$1.454E-2	&	1.500E-1	$_{4.700E-2}^{4.300E-2}$	&			\\
50-Sn-118	&	2.755E-1	$_{8.550E-2}^{7.577E-2}$	&	2.390E-1$\pm$2.247E-2	&	2.440E-1	$_{9.300E-2}^{1.310E-1}$	&			\\
50-Sn-119	&	1.431E-1	$_{2.387E-2}^{2.089E-2}$	&	9.700E-2$\pm$2.056E-2	&	1.840E-1	$_{6.900E-2}^{6.300E-2}$	&			\\
50-Sn-120	&	2.015E-1	$_{1.402E-1}^{1.223E-1}$	&	1.800E-1$\pm$1.710E-2	&	2.140E-1	$_{1.506E-1}^{1.980E-1}$	&			\\
51-Sb-121	&	1.121E-1	$_{9.715E-3}^{8.302E-3}$	&	1.048E-1$\pm$1.907E-2	&	8.360E-2	$_{2.580E-2}^{2.940E-2}$	&			\\
52-Te-122	&	-2.259E-2	$_{2.102E-2}^{1.794E-2}$	&	-6.324E-4$\pm$-6.324E-5	&							&	$s$-process only		\\
50-Sn-122	&	1.670E-1						&	1.616E-1$\pm$4.315E-2	&	1.520E-1	$_{1.520E-1}^{2.800E-2}$	&	$r$-process only		\\
52-Te-123	&	1.411E-3	$_{6.056E-3}^{5.148E-3}$	&	-1.241E-3$\pm$-1.241E-4	&							&	$s$-process only		\\
51-Sb-123	&	1.340E-1						&	1.298E-1$\pm$2.363E-2	&	1.130E-1	$_{2.050E-2}^{1.800E-2}$	&	$r$-process only		\\
52-Te-124	&	-2.131E-2	$_{3.603E-2}^{3.060E-2}$	&	-7.099E-3$\pm$-7.170E-4	&							&	$s$-process only		\\
50-Sn-124	&	2.090E-1						&		&	2.200E-1	$_{2.500E-2}^{2.200E-2}$	&	$r$-process only		\\
52-Te-125	&	2.567E-1	$_{1.144E-2}^{9.656E-3}$	&	2.579E-1$\pm$2.579E-2	&	2.560E-1	$_{3.900E-2}^{3.900E-2}$	&			\\
52-Te-126	&	4.812E-1	$_{5.960E-2}^{5.022E-2}$	&	4.800E-1$\pm$4.848E-2	&	4.920E-1	$_{1.110E-1}^{1.090E-1}$	&			\\
53-I-127	&	1.057E+0	$_{6.351E-3}^{5.297E-3}$	&	8.450E-1$\pm$1.817E-1	&	8.480E-1	$_{1.850E-1}^{1.820E-1}$	&			\\
54-Xe-128	&	1.177E-2	$_{1.619E-2}^{1.348E-2}$	&	4.500E-3$\pm$1.697E-3	&							&	$s$-process only		\\
52-Te-128	&	1.486E+0						&	1.526E+0$\pm$1.587E-1	&	1.470E+0	$_{1.800E-1}^{1.500E-1}$	&	$r$-process only		\\
54-Xe-129	&	1.426E+0	$_{1.079E-2}^{8.966E-3}$	&	1.234E+0$\pm$2.691E-1	&	1.080E+0	$_{2.290E-1}^{2.300E-1}$	&			\\
54-Xe-130	&	3.733E-2	$_{2.967E-2}^{2.460E-2}$	&	1.300E-2$\pm$4.472E-3	&							&	$s$-process only		\\
52-Te-130	&	1.585E+0						&		&	1.580E+0	$_{1.600E-1}^{1.600E-1}$	&	$r$-process only		\\
54-Xe-131	&	1.092E+0	$_{1.447E-2}^{1.193E-2}$	&	9.461E-1$\pm$2.536E-1	&	8.220E-1	$_{1.950E-1}^{1.880E-1}$	&			\\
54-Xe-132	&	8.091E-1	$_{9.283E-2}^{7.634E-2}$	&	7.480E-1$\pm$1.616E-1	&	6.530E-1	$_{2.640E-1}^{2.730E-1}$	&			\\
55-Cs-133	&	3.108E-1	$_{8.941E-3}^{7.232E-3}$	&	3.089E-1$\pm$2.162E-2	&	3.090E-1	$_{2.600E-2}^{3.200E-2}$	&			\\
56-Ba-134	&	-1.434E-2	$_{1.819E-2}^{1.469E-2}$	&	-6.322E-2$\pm$-4.489E-3	&							&	$s$-process only		\\
54-Xe-134	&	5.270E-1						&	4.507E-1$\pm$9.781E-2	&	3.850E-1	$_{1.540E-1}^{9.200E-2}$	&	$r$-process only		\\
56-Ba-135	&	2.384E-1	$_{8.434E-3}^{6.784E-3}$	&	2.155E-1$\pm$1.530E-2	&	2.480E-1	$_{2.480E-1}^{2.400E-2}$	&			\\
56-Ba-136	&	-3.803E-2	$_{5.797E-2}^{4.656E-2}$	&	-1.306E-1$\pm$-9.404E-3	&							&	$s$-process only		\\
54-Xe-136	&	4.290E-1						&		&	3.300E-1	$_{7.000E-2}^{6.600E-2}$	&	$r$-process only		\\
56-Ba-137	&	5.703E-2	$_{6.661E-2}^{5.289E-2}$	&	1.370E-1$\pm$1.000E-2	&	1.700E-1	$_{1.700E-1}^{1.260E-1}$	&			\\
56-Ba-138	&	-3.439E+0	$_{9.999E-1}^{7.834E-1}$	&	-5.796E-1$\pm$-3.767E-2	&	2.140E-1	$_{2.140E-1}^{7.860E-1}$	&			\\
57-La-139	&	7.730E-2	$_{5.975E-2}^{4.163E-2}$	&	7.700E-2$\pm$5.621E-3	&	1.570E-1	$_{1.387E-1}^{9.100E-2}$	&			\\
58-Ce-140	&	-5.995E-1	$_{2.604E-1}^{1.778E-1}$	&	3.200E-2$\pm$1.280E-3	&	1.610E-1	$_{1.610E-1}^{1.960E-1}$	&			\\
59-Pr-141	&	8.802E-2	$_{1.366E-2}^{8.675E-3}$	&	7.370E-2$\pm$1.990E-3	&	1.100E-1	$_{5.550E-2}^{2.600E-2}$	&			\\
60-Nd-142	&	-3.542E-2	$_{4.344E-2}^{2.739E-2}$	&	-2.700E-2$\pm$-6.480E-4	&							&	$s$-process only		\\
58-Ce-142	&	1.310E-1						&	1.143E-1$\pm$4.458E-3	&	6.600E-2	$_{6.600E-2}^{6.500E-2}$	&	$r$-process only		\\
60-Nd-143	&	6.871E-2	$_{5.632E-3}^{3.477E-3}$	&	6.250E-2$\pm$1.188E-3	&	7.060E-2	$_{1.800E-2}^{1.050E-2}$	&			\\
60-Nd-144	&	9.517E-2	$_{1.773E-2}^{1.091E-2}$	&	8.900E-2$\pm$1.958E-3	&	9.980E-2	$_{4.160E-2}^{2.420E-2}$	&			\\
60-Nd-145	&	5.653E-2	$_{3.047E-3}^{1.856E-3}$	&	4.670E-2$\pm$7.939E-4	&	5.400E-2	$_{8.400E-3}^{7.100E-3}$	&			\\
60-Nd-146	&	6.883E-2	$_{1.290E-2}^{7.846E-3}$	&	5.080E-2$\pm$8.636E-4	&	5.330E-2	$_{3.880E-2}^{1.780E-2}$	&			\\
60-Nd-147/ 61-Pm-147/ 62-Sm-147	&	3.327E-2$_{1.279E-3}^{7.720E-4}$$^{\dagger}$	&	3.099E-2$\pm$5.268E-4	&	3.340E-2	$_{1.780E-2}^{1.300E-3}$	&	$s$-process branching;	$\beta$-, T$_{1/2}$=10.98 d; $\beta$-, T$_{1/2}$=2.6234 y; $\alpha$, T$_{1/2}$=1.060x10$^{11}$ y	\\
62-Sm-148	&	-4.247E-4	$_{5.037E-3}^{3.037E-3}$	&	-5.256E-4$\pm$-8.410E-6	&							&	$s$-process only		\\
60-Nd-148	&	4.900E-2						&	4.393E-2$\pm$7.907E-4	&	4.210E-2	$_{2.000E-2}^{1.010E-2}$	&	$r$-process only		\\
62-Sm-149	&	3.287E-2	$_{6.851E-4}^{4.118E-4}$	&	3.111E-2$\pm$4.978E-4	&	3.230E-2	$_{4.500E-3}^{5.000E-4}$	&			\\
62-Sm-150	&	2.621E-3	$_{2.881E-3}^{1.730E-3}$	&	0.000E+0$\pm$0.000E+0	&							&	$s$-process only		\\
60-Nd-150	&	4.800E-2						&		&	4.900E-2	$_{3.100E-3}^{2.500E-3}$	&	$r$-process only		\\
62-Sm-151/ 63-Eu-151	&	-2.538E-3$_{4.209E-4}^{2.524E-4}$$^{\dagger}$	&	4.221E-2$\pm$1.815E-3	&	4.520E-2	$_{1.850E-2}^{3.000E-3}$	&	$s$-process branching;	$\beta$-, T$_{1/2}$=90 y	\\
62-Sm-152	&	5.511E-2	$_{2.636E-3}^{1.580E-3}$$^{\dagger}$	&	5.340E-2$\pm$8.544E-4	&	5.710E-2	$_{7.300E-3}^{5.100E-3}$	&	$s$-process branching		\\
63-Eu-153	&	4.868E-2	$_{4.511E-4}^{2.700E-4}$$^{\dagger}$	&	4.776E-2$\pm$4.824E-3	&	4.950E-2	$_{3.500E-3}^{3.100E-3}$	&	$s$-process branching		\\
64-Gd-154	&	1.980E-4	$_{1.262E-3}^{7.549E-4}$	&	9.300E-4$\pm$1.674E-5	&							&	$s$-process only		\\
62-Sm-154	&	6.000E-2						&	5.828E-2$\pm$3.847E-3	&	5.950E-2	$_{9.000E-3}^{1.400E-3}$	&	$r$-process only		\\
64-Gd-155	&	5.054E-2	$_{4.580E-4}^{2.738E-4}$	&	4.530E-2$\pm$8.154E-4	&	4.680E-2	$_{1.040E-2}^{3.200E-3}$	&			\\
64-Gd-156	&	6.157E-2	$_{1.998E-3}^{1.194E-3}$	&	5.480E-2$\pm$8.768E-4	&	5.790E-2	$_{7.800E-3}^{5.500E-3}$	&			\\
64-Gd-157	&	5.116E-2	$_{8.531E-4}^{5.093E-4}$	&	4.584E-2$\pm$8.251E-4	&	4.710E-2	$_{4.200E-3}^{3.700E-3}$	&			\\
64-Gd-158	&	6.610E-2	$_{3.872E-3}^{2.310E-3}$	&	5.800E-2$\pm$9.280E-4	&	6.140E-2	$_{1.170E-2}^{8.000E-3}$	&			\\
65-Tb-159	&	5.998E-2	$_{5.683E-4}^{3.383E-4}$	&	5.525E-2$\pm$3.315E-3	&	6.010E-2	$_{8.400E-3}^{7.100E-3}$	&			\\
66-Dy-160	&	8.963E-4	$_{1.414E-3}^{8.414E-4}$	&	8.800E-4$\pm$1.672E-5	&							&	$s$-process only		\\
64-Gd-160	&	7.870E-2						&	7.208E-2$\pm$6.992E-3	&	7.410E-2	$_{8.600E-3}^{4.600E-3}$	&	$r$-process only		\\
66-Dy-161	&	7.258E-2	$_{6.016E-4}^{3.577E-4}$	&	7.062E-2$\pm$1.201E-3	&	7.410E-2	$_{5.700E-3}^{4.000E-4}$	&			\\
66-Dy-162	&	8.729E-2	$_{2.580E-3}^{1.533E-3}$	&	8.440E-2$\pm$1.350E-3	&	9.000E-2	$_{1.050E-2}^{1.700E-3}$	&			\\
66-Dy-163	&	9.404E-2	$_{1.075E-3}^{6.382E-4}$	&	9.335E-2$\pm$1.867E-3	&	9.720E-2	$_{8.200E-3}^{8.000E-4}$	&			\\
66-Dy-164	&	8.187E-2	$_{5.366E-3}^{3.182E-3}$	&	9.530E-2$\pm$1.811E-3	&	1.030E-1	$_{2.030E-2}^{1.000E-3}$	&			\\
67-Ho-165	&	8.580E-2	$_{8.663E-4}^{5.122E-4}$	&	8.308E-2$\pm$4.652E-3	&	8.390E-2	$_{1.110E-2}^{1.020E-2}$	&			\\
68-Er-166	&	7.812E-2	$_{1.646E-3}^{9.727E-4}$	&	7.090E-2$\pm$7.161E-3	&	7.530E-2	$_{6.200E-3}^{8.000E-3}$	&			\\
68-Er-167	&	5.558E-2	$_{7.364E-4}^{4.347E-4}$	&	5.237E-2$\pm$5.289E-3	&	5.460E-2	$_{5.100E-3}^{4.000E-3}$	&			\\
68-Er-168	&	4.833E-2	$_{3.780E-3}^{2.230E-3}$	&	4.550E-2$\pm$5.506E-3	&	5.060E-2	$_{8.600E-3}^{6.400E-3}$	&			\\
69-Tm-169	&	3.430E-2	$_{1.051E-3}^{6.185E-4}$	&	3.102E-2$\pm$1.706E-3	&	3.400E-2	$_{9.000E-3}^{5.100E-3}$	&			\\
70-Yb-170	&	-1.136E-3	$_{1.458E-3}^{8.579E-4}$	&	1.000E-3$\pm$4.200E-5	&							&	$s$-process only		\\
68-Er-170	&	3.900E-2						&	3.521E-2$\pm$5.246E-3	&	3.690E-2	$_{8.600E-3}^{3.800E-3}$	&	$r$-process only		\\
70-Yb-171	&	3.072E-2	$_{8.989E-4}^{5.285E-4}$	&	2.933E-2$\pm$1.144E-3	&	2.970E-2	$_{1.900E-2}^{2.900E-3}$	&			\\
70-Yb-172	&	3.657E-2	$_{3.177E-3}^{1.867E-3}$	&	3.660E-2$\pm$3.074E-3	&	3.810E-2	$_{5.800E-3}^{5.100E-3}$	&			\\
70-Yb-173	&	3.264E-2	$_{1.446E-3}^{8.483E-4}$	&	3.108E-2$\pm$2.642E-3	&	3.160E-2	$_{5.000E-3}^{3.700E-3}$	&			\\
70-Yb-174	&	4.004E-2	$_{7.030E-3}^{4.119E-3}$	&	3.610E-2$\pm$3.321E-3	&	3.910E-2	$_{1.620E-2}^{1.240E-2}$	&			\\
71-Lu-175	&	3.201E-2	$_{8.344E-4}^{4.869E-4}$	&	2.979E-2$\pm$1.192E-3	&	3.050E-2	$_{1.490E-2}^{6.900E-3}$	&			\\
71-Lu-176/ 72-Hf-176	&	-3.149E-3$_{7.113E-4}^{4.148E-4}$$^{\dagger}$	&	-8.549E-4$\pm$-3.505E-5	&							& $s$-process branching;	$s$-process only;	$\beta$-, T$_{1/2}$=3.76x10$^{10}$ y	\\
70-Yb-176	&	3.330E-2						&	3.040E-2$\pm$3.040E-3	&	2.920E-2	$_{1.150E-2}^{4.200E-3}$	&	$r$-process only		\\
72-Hf-177	&	2.451E-2	$_{7.513E-4}^{4.379E-4}$	&	2.356E-2$\pm$1.154E-3	&	2.380E-2	$_{5.200E-3}^{2.500E-3}$	&			\\
72-Hf-178	&	2.102E-2	$_{3.597E-3}^{2.095E-3}$	&	2.010E-2$\pm$7.437E-4	&	1.920E-2	$_{9.200E-3}^{4.400E-3}$	&			\\
72-Hf-179	&	1.446E-2	$_{1.129E-3}^{6.562E-4}$	&	1.420E-2$\pm$5.112E-4	&	1.380E-2	$_{2.900E-3}^{2.200E-3}$	&			\\
72-Hf-180	&	2.411E-2	$_{5.128E-3}^{2.978E-3}$	&	1.700E-2$\pm$5.780E-4	&	1.450E-2	$_{1.450E-2}^{6.900E-3}$	&			\\
73-Ta-181	&	1.353E-2	$_{1.254E-3}^{7.260E-4}$	&	1.217E-2$\pm$3.286E-4	&	1.060E-2	$_{6.400E-3}^{3.800E-3}$	&			\\
74-W-182	&	1.419E-2	$_{3.711E-3}^{2.146E-3}$	&	2.150E-2$\pm$1.548E-3	&	1.360E-2	$_{1.360E-2}^{7.900E-3}$	&			\\
74-W-183	&	8.727E-3	$_{1.827E-3}^{1.054E-3}$	&	6.600E-3$\pm$4.752E-4	&	6.500E-3	$_{6.500E-3}^{3.500E-3}$	&			\\
74-W-184	&	1.723E-2	$_{4.162E-3}^{2.398E-3}$	&	1.260E-2$\pm$9.198E-4	&	1.060E-2	$_{1.060E-2}^{7.300E-3}$	&			\\
75-Re-185	&	1.537E-2	$_{8.971E-4}^{5.154E-4}$	&	1.339E-2$\pm$1.366E-3	&	1.510E-2	$_{4.100E-3}^{2.500E-3}$	&			\\
76-Os-186	&	-2.681E-3	$_{2.268E-3}^{1.302E-3}$	&	1.000E-4$\pm$7.400E-6	&							&	$s$-process only		\\
74-W-186	&	3.900E-2						&	2.959E-2$\pm$2.071E-3	&	2.450E-2	$_{1.720E-2}^{9.200E-3}$	&	$r$-process only		\\
76-Os-187	&	3.447E-3	$_{8.675E-4}^{4.973E-4}$	&	3.080E-3$\pm$2.187E-4	&	3.180E-2	$_{4.800E-3}^{4.100E-3}$	&	contribution of $^{197}$Re, T$_{1/2}$=4.33x10$^{10}$ y		\\
76-Os-188	&	7.356E-2	$_{2.835E-3}^{1.624E-3}$	&	7.440E-2$\pm$5.431E-3	&	7.080E-2	$_{7.500E-3}^{7.300E-3}$	&			\\
76-Os-189	&	1.044E-1	$_{9.420E-4}^{5.386E-4}$	&	1.038E-1$\pm$7.781E-3	&	1.030E-1	$_{6.900E-3}^{6.000E-3}$	&			\\
76-Os-190	&	1.621E-1	$_{2.852E-3}^{1.630E-3}$	&	1.576E-1$\pm$2.600E-2	&	1.520E-1	$_{1.500E-2}^{1.600E-2}$	&			\\
77-Ir-191	&	2.453E-1	$_{7.945E-4}^{4.530E-4}$	&	2.424E-1$\pm$1.915E-2	&	2.290E-1	$_{8.000E-3}^{8.000E-3}$	&			\\
78-Pt-192	&	-4.724E-3	$_{2.483E-3}^{1.415E-3}$	&	-5.355E-3$\pm$-2.704E-3	&							&	$s$-process only		\\
76-Os-192	&	2.780E-1						&	2.760E-1$\pm$4.360E-2	&	2.730E-1	$_{2.100E-2}^{1.600E-2}$	&	$r$-process only		\\
78-Pt-193/ 77-Ir-193	&	4.137E-1$_{1.225E-3}^{6.970E-4}$$^{\dagger}$	&	4.079E-1$\pm$3.222E-2	&	3.880E-1	$_{1.400E-2}^{1.400E-2}$	& $s$-process branching;		EC,T$_{1/2}$=50 y	\\
78-Pt-194	&	4.014E-1	$_{3.135E-3}^{1.782E-3}$	&	4.223E-1$\pm$2.133E-1	&	4.210E-1	$_{5.900E-2}^{4.900E-2}$	&			\\
78-Pt-195	&	4.237E-1	$_{1.236E-3}^{7.005E-4}$	&	4.474E-1$\pm$2.259E-1	&	4.450E-1	$_{5.100E-2}^{4.800E-2}$	&			\\
78-Pt-196	&	2.929E-1	$_{4.914E-3}^{2.784E-3}$	&	3.085E-1$\pm$4.257E-2	&	3.020E-1	$_{4.600E-2}^{4.500E-2}$	&			\\
79-Au-197	&	1.857E-1	$_{1.574E-3}^{8.884E-4}$	&	1.773E-1$\pm$2.677E-2	&	1.910E-1	$_{1.200E-2}^{1.300E-2}$	&			\\
80-Hg-198	&	1.051E-2	$_{6.007E-3}^{3.386E-3}$	&	2.700E-3$\pm$3.996E-4	&							&	$s$-process only		\\
78-Pt-198	&	9.100E-2						&	9.624E-2$\pm$1.165E-2	&	9.500E-2	$_{1.450E-2}^{1.000E-2}$	&	$r$-process only		\\
80-Hg-199	&	6.263E-2	$_{2.435E-3}^{1.366E-3}$	&	4.300E-2$\pm$5.805E-3	&	5.070E-2	$_{1.500E-2}^{1.750E-2}$	&			\\
80-Hg-200	&	6.233E-2	$_{7.407E-3}^{4.147E-3}$	&	3.350E-2$\pm$5.327E-3	&	3.340E-2	$_{2.730E-2}^{3.060E-2}$	&			\\
80-Hg-201	&	3.868E-2	$_{3.623E-3}^{2.017E-3}$	&	2.550E-2$\pm$3.341E-3	&	2.650E-2	$_{1.540E-2}^{1.610E-2}$	&			\\
80-Hg-202	&	7.395E-2	$_{1.072E-2}^{5.951E-3}$	&	3.760E-2$\pm$5.452E-3	&	2.570E-2	$_{2.570E-2}^{4.200E-2}$	&			\\
81-Tl-203	&	2.352E-2	$_{5.199E-3}^{2.860E-3}$	&	1.730E-2$\pm$1.972E-3	&	3.300E-3	$_{3.300E-3}^{2.380E-2}$	&			\\
82-Pb-204	&	-4.481E-3	$_{1.204E-2}^{6.593E-3}$	&	1.280E-2$\pm$1.267E-3	&							&	$s$-process only		\\
80-Hg-204	&	3.100E-2						&	2.295E-2$\pm$3.511E-3	&	2.660E-2	$_{9.500E-3}^{6.400E-3}$	&	$r$-process only		\\
82-Pb-205/ 81-Tl-205	&	-2.318E-2	$_{3.971E-3}^{2.153E-3}$	&	5.390E-2$\pm$6.468E-3	&	4.970E-2	$_{4.970E-2}^{6.530E-2}$	&	$^{237}$Np, $^{243}$Am-series;	EC, T$_{1/2}$=1.73x10$^{7}$ y	 \\
82-Pb-206	&	2.588E-1	$_{6.093E-2}^{3.291E-2}$	&	4.090E-1$\pm$4.335E-2	&	1.970E-1	$_{1.606E-1}^{1.820E-1}$	&	$^{238}U$-series		\\
82-Pb-207	&	2.014E-1	$_{8.325E-2}^{4.298E-2}$	&	4.540E-1$\pm$4.268E-2	&	1.420E-1	$_{1.420E-1}^{2.910E-1}$	&	$^{235}$U, $^{239}$Pu, $^{247}$Cm, $^{231}$Pa-series		\\
82-Pb-208	&	-2.477E+0	$_{7.844E-1}^{3.794E-1}$	&	1.649E+0$\pm$1.385E-1	&	3.000E-4	$_{3.000E-4}^{1.780E+0}$	&	$^{232}$Th-series		\\
83-Bi-209	&	-1.671E-2	$_{2.891E-2}^{1.160E-2}$	&	1.391E-1$\pm$1.822E-2	&	5.010E-2	$_{4.010E-2}^{1.139E-1}$	&			\\

\hline \hline
\end{longtable}
\end{footnotesize}







\end{document}